\documentclass[12pt]{article}
\usepackage{amssymb}
  
\catcode`@=11
\def\secteqno{\@addtoreset{equation}{section}%
\def\theequation{\thesection.\arabic{equation}}}
\catcode`@=12
\secteqno
\newcommand{\be}{\begin{equation}}
\newcommand{\ee}{\end{equation}}
\newcommand{\bea}{\begin{eqnarray}}
\newcommand{\eea}{\end{eqnarray}}
\newcommand{\s}{-}
\newcommand{\bref}[1]{(\ref{#1})}
\newcommand{\nn}{\nonumber}	   

\mathcode`\*="702A                  
\catcode163=13 \def\itm{\relax\ifmmode\to\else\itemize\fi}
\begin{document}
\thispagestyle{empty}
\hfill May 24, 2004

\hfill KEK-TH-956,
       Toho-CP-0475

\vskip 20mm
\begin{center}
{\Large\bf Wess-Zumino terms for AdS D-branes }
\vskip 6mm
\medskip
\vskip 10mm
{\large Machiko\ Hatsuda$^{\ast\dagger}$~and~Kiyoshi\ Kamimura$^\star$}

\parskip .15in
{\it $^\ast$Theory Division,\ High Energy Accelerator Research Organization (KEK),\\
\ Tsukuba,\ Ibaraki,\ 305-0801, Japan} \\{\it $^\dagger$Urawa University, Saitama \ 336-0974, Japan}\\
{\small e-mail:\ 
{mhatsuda@post.kek.jp}} \\
\parskip .35in
{\it 
$~^\star$ 
 Department of Physics, Toho University, Funabashi, 274-8510, Japan}\\
 {\small e-mail:\ 
 {kamimura@ph.sci.toho-u.ac.jp} }\\

\medskip
\end{center}
\vskip 10mm
\begin{abstract}
We show that Wess-Zumino terms for D-$p$ branes with $p>0$
in the Anti-de Sitter (AdS) space are given  
in terms of ``left-invariant" currents on   
the super-AdS group or the ``expanded" super-AdS group.
As a result there is no topological extension of the super-AdS algebra.
In the flat limit the global Lorentz rotational charges of the AdS space
turn out to be brane charges of the supertranslation algebra 
representing the BPS mass.
We also show that a D-instanton is described by the $GL(1)$ degree of freedom
in the Roiban-Siegel formalism based on the $GL(4{\mid}4)/[Sp(4)\times GL(1)]^2$ coset.
\end{abstract} 

\setcounter{page}{1}
\parskip=7pt
\newpage
\section{ Introduction}

Dynamics of supersymmetric extended objects 
play essential  roles in nonperturbative aspects of 
superstring theories. 
A supersymmetric $p$-brane  
is described by an action with a Wess-Zumino (WZ) term.
In contrast to that geometrical interpretation of 
flat $p$-branes are discussed well in 
\cite{AzTow1} and \cite{AzTow2}, 
the one for AdS $p$-branes has not been clarified.
The WZ term for a flat $p$-brane 
is characterized by the nontrivial element of
the $(p+2)$-th Chevalley-Eilenberg cohomology
of the super-translation group \cite{AzTow1},
and the superalgebra contains topological charges
originated from the WZ term
\cite{AzTow2}.
The superalgebra gives a BPS condition
with the BPS mass determined by the topological charge.

Chevalley and Eilenberg formulated cohomology theory of Lie algebras
\cite{CE}: let 
$C^q({\mathcal G},P)$ be a vector space of $q$-th cochain
for a representation $P$ of a Lie algebra ${\mathcal G}$ 
over a field of characteristic $0$. 
In our case $c\in C^q({\mathcal G},{\mathbb R})$ 
is a left-invariant (LI) differential 
$q$ form on the group ${\mathcal G}$ mapping to real space ${\mathbb R}$,
and then the cohomology group is defined as
$H^q({\mathcal G},{\mathbb R})=Z^q({\mathcal G},{\mathbb R})/B^q({\mathcal G},{\mathbb R})$
with $Z^q({\mathcal G},{\mathbb R})=\{c\in C^q({\mathcal G},{\mathbb R})\mid dc=0\}$
and 
$B^q({\mathcal G},{\mathbb R})=\{dc\mid c \in C^{q-1}({\mathcal G},{\mathbb R})\}$.
A super-$p$-brane in a flat space
is described by the $(p+1)$-form WZ term $c$, and 
is classified by the nontrivial class of the 
$(p+2)$-th CE cohomology of ${\mathcal G}=$ the ``supertranslation" group
\bea
dr=0,~~r= dc,~~r\in C^{p+2}({\mathcal G},{\mathbb R}),~~
c\notin C^{p+1}({\mathcal G},{\mathbb R})~~~.
\eea
The WZ term is not manifestly superinvariant but pseudo-invariant
so the supercharge is modified,
and this modification contributes to topological terms in the flat superalgebra.

Recently 
brane dynamics in AdS spaces have been examined widely 
following the early works by Metsaev and Tseytlin  \cite{MT,MTD3}. 
In these references the WZ terms of $p$-branes for $p=1,3$ are given 
as a $(p+2)$-dimensional integration of the 
closed $(p+2)$-forms.
On the other hand it was shown that in AdS spaces
the WZ terms for $p$-branes can be  
written in terms of the left-invariant (LI) currents
for $p=1$ cases \cite{Berk,RWS,AdSWZ,polyakov},
and for a $p=0$ case \cite{MHD0}.  
These WZ terms are manifestly superinvariant
so there is no modification of the supercharges,
therefore 
no topological charge in the superalgebras appears.
It was also shown that the topological brane charges in a flat 
superalgebra 
can be traced back into the super-AdS algebra through the limiting procedure.
 For a BPS 0-brane in AdS$_2\times$S$^2$ space 
the origin of the 0-brane charge is the Lorentz rotation charge of the AdS algebra \cite{MHD0}.
To clarify these issues for more general cases,
in this paper we focus on  D$p$-branes for $p=-1,1,3$ in the AdS$_5\times$S$^5$ space.

The closed 5-form for an AdS D3-brane is obtained in the reference \cite{MTD3}.
It is not simple enough to find the explicit local  form of the  WZ term
to examine the superalgebra and the brane charges.
In this paper we use ``expansion" technique.
It was originally introduced in
\cite{AdSWZ,HSexp}
as a generalization of 
the In$\ddot{\rm o}$n$\ddot{\rm u}$-Wigner (IW) contraction  \cite{IW},
and further developed in \cite{Azexp}.
The IW contraction relates the super-AdS algebra with the 
supertranslation algebra,
while the generalized IW contraction is useful to relate
the WZ term in the AdS space and the one in the flat space \cite{AdSWZ}.
This generalization is a contraction in which 
the next to leading terms in the expansion parameter, inverse of the AdS radius, are kept.
It was also shown that the expansion truncated 
at certain order gives another closed algebra \cite{HSexp}.
In this paper we will show that 
this ``expansion" procedure without truncation
is useful to analyze the WZ term.

The organization of this paper is the following.
In the next section
we will review the Roiban-Siegel formulation
\cite{RWS}
which makes these computation possible. 
In the section 3 after reviewing the closed 3-form and 
its 2-form WZ term
for an AdS superstring, we examine the ones for an AdS D-string.
By taking care of $GL(1)$ degree of freedom carefully,
the field strength for an AdS D-instanton is obtained.
In the section 4, 
we first examine the closed 5-form  and the 4-form WZ term 
for an bosonic D3-brane.
We will show that it can not be written in terms of 
LI currents of AdS$_5$ algebra.
We analyze it by using the  ``expansion" procedure. 
It turns out that the 4-form potential can be written 
in terms of the ``expanded" currents of the AdS$_5$ LI currents.
Next we examine the ones for an super D3-brane.
In the section 5 the flat limit is discussed.
We will discuss that there is no topological extension of the super-AdS algebra
by examining the super-variation of the WZ term concretely.
We will also discuss the origin of the topological terms of the 
flat superalgebra in the super-AdS algebra.

\section{ Notations}

The non-linear realization of the super AdS$_5\times$S$^5$ group  has been 
discussed in \cite{MT,MTD3} using 
\bea
\displaystyle\frac{PSU(2,2{\mid}4)}{SO(5,1)\otimes SO(6)}~~~.
\eea
In an alternative formulation of it Roiban and Siegel used the coset \cite{RWS}  
\bea
{\mathcal G}=\displaystyle\frac{GL(4{\mid}4)}{\left[Sp(4)\otimes GL(1)\right]^2}~~~\sim~~~
\displaystyle\frac{PSU(2,2{\mid}4)}{SO(5,1)\otimes SO(6)}~~~.
\eea
It requires a Wick rotation for the matrix valued coordinate of ${\mathcal G}$ 
to be the real coordinates, 
and introduces scaling factors of $\left[GL(1)\right]^2$.  
This formulation
made computations of the equations of motion and the symmetry algebras
much easier \cite{HKAdS}. 
We follow the notations of
\cite{RWS} and \cite{HKAdS} in this paper. 

An element of the coset $Z_M{}^A$ is transformed under 
a global $GL(4{\mid}4)$ group element $g$ with the indices $M=m,\bar{m}$, 
$(m,\bar{m}=1,\cdots, 4)$ as
\bea
Z_M{}^A~\rightarrow~g_M{}^N~Z_N{}^B ~h_B{}^A~~~
\eea
where  $h$ is a local  $\left[Sp(4)\otimes GL(1)\right]^2$ group element 
with $A=a,\bar{a}$, $(a,\bar{a}=1,\cdots, 4)$.
The LI 1-form current is given by
\bea
J_A{}^B=Z_A{}^MdZ_M{}^B~~~
\label{LIJ}\eea 
and transforms as 
\bea
J&\to&h^{-1}~J~h~+~h^{-1}~d~h.
\label{htrans}
\eea
The bosonic components are decomposed as 
\bea
J^{ab}&=&J^{\langle ab\rangle}+J^{(ab)}-\frac{1}{4}C^{ab}{\rm tr}J\equiv
 \langle{\bf J}\rangle+({\bf J})-\frac{1}{4}~C~{\rm tr}{\bf J}\\
 J^{\bar{a}\bar{b}}&=&J^{\langle \bar{a}\bar{b}\rangle}+J^{(\bar{a}\bar{b})}
 -\frac{1}{4}C^{\bar{a}\bar{b}}{\rm tr}J\equiv
 \langle\bar{\bf J}\rangle+(\bar{\bf J})-\frac{1}{4}~C~{\rm tr}\bar{\bf J}\nn
\eea
The index $\langle ab\rangle$ denotes antisymmetric traceless index 
which can be 
expanded by
$(C\gamma_\alpha)^{ab}$,
 and the index $( ab)$ denotes symmetric index expanded by
$(C\gamma_{\alpha\beta})^{ab}$ with  the 5-dimensional gamma matrices
$\gamma_\alpha,~(\alpha=0,\cdots, 4)$. 
The $Sp(4)$ invariant antisymmetric  
metric  $C$, denoted by ``$\Omega$" in \cite{RWS},
is the charge conjugation matrix, with $C^{-1}=C^T=-C$,  
in this paper. Trace is taken as tr$M=M^a{}_a=M^{ab}C_{ba}$.
The fermionic components of $J$ in \bref{LIJ} will be denoted by small
characters;
\bea
J^{a\bar{b}}=j^{a\bar{b}}~~,~~J^{\bar{a}b}=\bar{j}^{\bar{a}b}~~~.
\eea

These LI 1-forms satisfy the Maurer-Cartan (MC) equations,
$-dJ_A{}^B=J_A{}^CJ_C{}^B$, more explicitly
\bea
\left\{\begin{array}{rcl}
-d\langle{\bf J}\rangle&=&({\bf J})\langle{\bf J}\rangle+
\langle{\bf J}\rangle({\bf J})+\langle j\bar{j}\rangle
=\{({\bf J}),\langle{\bf J}\rangle\}+\langle j\bar{j}\rangle
\\
-d({\bf J})&=&\langle{\bf J}\rangle\langle{\bf J}\rangle
+({\bf J})({\bf J})+( j\bar{j})
=\langle{\bf J}\rangle^2+({\bf J})^2+( j\bar{j})
\\
-dj&=&\langle{\bf J}\rangle j+({\bf J})j+
j\langle\bar{\bf J}\rangle+j(\bar{\bf J})-\frac{1}{4}({\rm Str} J) j
\\
-d\bar{j}&=&\langle\bar{\bf J}\rangle \bar{j}+(\bar{\bf J})\bar{j}+
\bar{j}\langle{\bf J}\rangle+\bar{j}({\bf J})+\frac{1}{4}({\rm Str} J) \bar{j}
\end{array}\right.\label{MCGL44}
\eea
and  similar expressions for barred sectors. 
It is noted that we use a notation in which the exterior derivative 
``$d$" commutes with both Grassman even and odd components of ``$Z_M{}^A$".
\par\vskip 6mm
\section{ 2-form WZ terms}

\subsection{ Superstring}
The $\kappa$-symmetry of a superstring requires the 2-form WZ term.
It was shown in \cite{RWS} that
the WZ term for an AdS-string is given as 
\bea
{\cal L}_{WZ,{\rm AdS},{\rm F1}}&=&B_{[2]}^{\rm NS}=B_{[2]-}\label{31}
\eea
with
\bea
B_{[2]\pm}&=&\frac{1}{2}\left\{
E^{1/2} J^{a\bar{b}} J_{a\bar{b}}\pm
E^{-1/2}J^{\bar{a}b} J_{\bar{a}{b}}\right\}=
\frac{1}{2}{\rm tr}\left\{E^{1/2}jj\pm
E^{-1/2} \bar{j}\bar{j}
\right\}~~~\label{Bpm}
\eea
and it has the correct flat limit. 
The invariance under the local $GL(1)$ scaling transformations 
requires introducing $E={\rm Sdet} Z_M{}^A$ factors. 
In the gauge $E=1$ the coset 
${\mathcal G}=GL(4{\mid}4)/[Sp(4){\otimes}GL(1)]^2$ reduces into
$PSL(4{\mid}4)/[Sp(4)]^2$.
The closed three form is given by
\bea
dB_{[2]}^{\rm NS}=H_{[3]}^{\rm NS}~~,~~H_{[3]}^{\rm NS}=H_{[3]+}\label{33}
\eea
with
\bea
H_{[3]\pm}&=&\langle{\bf J}\rangle^{ab}\left\{
E^{1/2}\langle J_a{}^{\bar{b}} J_{b\bar{b}}\rangle\pm
E^{-1/2}\langle J^{\bar{a}}{}_{a} J_{\bar{a}{b}}\rangle
\right\}-
\langle{\bf J}\rangle^{\bar{a}\bar{b}}\left\{E^{1/2}\langle J^a{}_{\bar{a}} J_{a\bar{b}}\rangle\pm
E^{-1/2}\langle J_{\bar{a}}{}^{a} J_{\bar{b}{a}}\rangle
\right\}\nn\\
&=&-{\rm tr}\left[\langle{\bf J}\rangle\left\{E^{1/2}\langle jj\rangle\pm
E^{-1/2}\langle \bar{j}\bar{j}\rangle
\right\}-
\langle\bar{\bf J}\rangle\left\{E^{1/2}\langle jj\rangle\pm
E^{-1/2}\langle \bar{j}\bar{j}\rangle
\right\}
\right]~~~\label{H3pmpm}
\eea
satisfying
\bea
d(B_{[2]\mp})=H_{[3]\pm}~~~.\label{H3p}
\eea
Then the 3-form field strengths in the super-AdS space \bref{H3p}
are trivial elements of the 3-rd CE cohomology of 
${\mathcal G}$, 
\bea
dH_{[3]}=0,~~H_{[3]}= dB_{[2]},~~H_{[3]}\in C^{3}({\mathcal G},{\mathbb R}),~~
B_{[2]}\in C^{2}({\mathcal G},{\mathbb R})~~~.\label{CEAdSF1}
\eea
Since these left invariant 2-forms are manifestly global superinvariant,
the supercharges 
do not contain additional shift from the WZ action. 
So there is no topological extension of the superalgebra for a AdS-superstring.
\par

\subsection{ D-string and D-instanton}

The WZ term for D$p$-brane is given as the integral of the closed and invariant $(p+2)$~-form, 
$d{\cal L}_{WZ}$,
\bea
d{\cal L}_{WZ}=e^{\cal F}{\cal R}=e^{dA}\hat{\cal R}~~,~~
{\cal F}=dA-B_{[2]}^{\rm NS}~~,~~
\hat{\cal R}=e^{-B_{[2]}^{\rm NS}}{\cal R}
\label{generalR}
\eea
where  $\hat{\cal R}$ satisfies $d\hat{\cal R}=0$ while
$d{\cal R}=H_{[3]}^{\rm NS}{\cal R}$. 
We comment  that 
the Dirac-Born-Infeld U(1) gauge field $A$ and $B_{[2]}^{\rm NS}$
are separately superinvariant, 
$\delta_\epsilon A=\delta_\epsilon B_{[2]}^{\rm NS}=0$,
 in our formulation.
This is possible by the existence of total derivative bilinear term,
 $d ~(\theta d\bar{\theta})$, in $B_{[2]}^{\rm NS}$
which can not be introduced in the conventional 2-form 
$\displaystyle\int H_{[3]}^{\rm NS}$ 
as discussed in \cite{AdSWZ}.

Recently it is observed that $PSU(2,2{\mid}4)$
is extended to $SU(2,2{\mid}4)$ by an external automorphism $U(1)_B$
under the circumstance of the 
existence of BPS multiplets in the ${\cal N}=4$ SYM \cite{beisert}.
In our description 
the $E=1$ gauge fixing of $GL(1)$ plays the similar role, so
we also keep the $GL(1)$ field unfixed.
In the reference \cite{RWS}  the auxiliary 
$GL(1)$ field, $E$, is introduced to preserve the local $GL(1)$ invariance
of the action \bref{31} and \bref{Bpm}, and the resultant three-form field strength
includes $E$ \bref{33} and \bref{H3pmpm}
.
For the R/R two-form
we instead break the local $GL(1)$ invariance 
to take into account the BPS effect as observed above.
The 3-form field strength is determined
in such a way that it does not depend on $E$,
\bea
\hat{\cal C}_{[2]}=
\check{B}_{[2]+}~~,~~ B_{[2]}^{\rm NS}=
\check{B}_{[2]-}\label{C2RR}~~,~~
\check{B}_{[2]\pm}=
\frac{1}{2}{\rm tr}\left\{jj\pm
 \bar{j}\bar{j}
\right\}~~~\label{checkBpm}
\eea
satisfying
\bea
d\check{B}_{[2]\pm}=\check{H}_{[3]\mp}+\frac{1}{2}{\rm Str}J~\check{B}_{[2]\mp}
~~,~~\check{H}_{[3]\pm}=
-{\rm tr}\left[\langle{\bf J}\rangle\left\{\langle jj\rangle\pm
\langle \bar{j}\bar{j}\rangle
\right\}-
\langle\bar{\bf J}\rangle\left\{\langle jj\rangle\pm
\langle \bar{j}\bar{j}\rangle
\right\}
\right]~~\nn~\\\label{checkH3pmpm}~~~.
\eea
Then the exterior derivative of the R/R two-form becomes
\bea 
\hat{\cal R}_{[3]}
\equiv
d\hat{\cal C}_{[2]}
=\check{H}_{[3]-}
+\frac{1}{2}{\rm Str}J~\check{B}_{[2]-}~~~.\label{r3r3}
\eea
On the other hand it is also written as
\bea
\hat{\cal R}_{[3]}
={\cal R}_{[3]}-B_{[2]}^{\rm NS}~{\cal R}_{[1]}~
~~,~
\left\{\begin{array}{ccl}
{\cal R}_{[3]}&=&\check{H}_{[3]-}\\
{\cal R}_{[1]}&=&-\frac{1}{2}{\rm Str} J\end{array}\right.~~~
\eea
from \bref{generalR}. 
If we identify ${\cal R}_{[1]}$ 
as the field strength of a D-instanton,
then the WZ term is given as 
\bea
{\cal L}_{WZ,{\rm AdS},{\rm D~instanton}}={\cal C}_{[0]}
~~,~~{\cal C}_{[0]}
=\frac{1}{2}{\rm ln} E~~
\eea 
where Str$J=-d$~ln$E=0$ is used.
With this degree of freedom,
 the WZ term for an AdS-D-string is given by
\bea
{\cal L}_{WZ,{\rm AdS},{\rm D1}}=\hat{\cal C}_{[2]}+dA~{\cal C}_{[0]}
~~.
\eea
The WZ terms for AdS D-string and AdS D-instanton are written in terms of the 
LI 1-form currents or $E={\rm Sdet} Z_M{}^A$ which are manifestly superinvariant.
So there is no contribution to the supercharges from these WZ terms,
then no topological term is produced in the AdS superalgebra.

\par\vskip 6mm
\section{ 4-form WZ term}

In the AdS$_5$ space the closed 5-form has a contribution from 
the 5-form RR flux \cite{MTD3}.
At first we will concentrate on this bosonic 5-form RR flux effect
and then we will extend it to the supersymmetric case.

\subsection{ Bosonic D3-brane}

We begin with the bosonic AdS$_5$ background where
${\bf z}_m{}^a$ is an element of
the bosonic coset $G=GL(4)/[Sp(4)\otimes GL(1)]$  
and the LI current is ${\bf j}_a{}^b={\bf z}_a{}^md{\bf z}_m{}^b$.
These currents are basis of the $q$-forms on
$G$.
They satisfy the following MC equations
\bea
\left\{
\begin{array}{rcl}
-d\langle{\bf j} \rangle&=&\{({\bf j}),\langle{\bf j}\rangle\}\\
-d({\bf j})&=&({\bf j})^2+\langle{\bf j}\rangle^2
\end{array}
\right.\label{mcbose}
\eea
Using them the closed and invariant ``bosonic"  5-form is 
\bea
\hat{R}_{[5]}={\rm tr} ~\langle{\bf j}\rangle^5~~~.\label{r5boson}
\label{r5}
\eea
The corresponding 4-form potential
\bea
\hat{R}_{[5]}=d\hat{C}_{[4]}~~,~~\hat{C}_{[4]}=\int \hat{R}_{[5]}\label{R5dC4}
\eea
can be evaluated explicitly as follows.
We take the local Lorentz $Sp(4)$ and the $GL(1)$ gauge choice as a ``unitary gauge"
\bea
&&{\bf z}_m{}^a=\frac{1}{\sqrt{1-x^2}}{\bf 1}_m{}^b\left\{{\bf 1}+x^\alpha\gamma_\alpha\right\}_b{}^a,~
{\bf z}_a{}^m=\displaystyle\frac{1}{\sqrt{1-x^2}} 
 \left\{{\bf 1}-x^\alpha\gamma_\alpha\right\}_a{}^b{\bf 1}_b{}^m~,
\label{unitary}\label{zxz}
\eea
where $x^\alpha (\alpha=0,\cdots, 4)$ are  the AdS$_5$ coordinates.  
The LI currents, ${\bf j}_a{}^b={\bf z}_a{}^md{\bf z}_m{}^b
$, are
\bea
&\langle{\bf j}_{\rm unitary}\rangle_a{}^b=\displaystyle\frac{dx^\alpha}{1-x^2} 
(\gamma_\alpha)_a{}^b~,~
({\bf j}_{\rm unitary})_a{}^b=-\displaystyle\frac{x^\alpha dx^\beta}{1-x^2} 
 (\gamma_{\alpha\beta})_a{}^b~,~
{\rm tr}~{\bf j}_{\rm unitary}=d\left\{{\rm ln} (1-x^2)\right\}&\nn\\&&\label{unij}
\eea
and they satisfy the MC equations \bref{mcbose}.
Using the Cartan's homotopy formula \cite{AI},
where $~x_t=tx~$, $~k_t x_t\equiv 0~$ and  $~k_t dx_t\equiv x~$,
the ``bosonic" 4-form potential is given as
\bea
\hat{C}_{[4]}&=&\displaystyle\int_0^1 dt~ k_t~ R_{[5]}(x_t)\nn\\
&=&\displaystyle\int_0^1 dt \frac{5t^4}{(1-t^2x^2)^5} x^\alpha dx^\beta dx^\gamma dx^\delta dx^\epsilon
\varepsilon_{\alpha\beta\gamma\delta\epsilon}\nn\\
&=&f(x)~x^\alpha dx^\beta dx^\gamma dx^\delta dx^\epsilon
\varepsilon_{\alpha\beta\gamma\delta\epsilon}
\eea
with
\bea
f(x)&=&\displaystyle\int_0^1 dt \frac{5t^4}{(1-t^2x^2)^5}\nn\\
&=&\displaystyle\frac{5}{128 |x|^5(1-|x|^2)^4}
\left[-3|x|(1+|x|^6)+11|x|^3(1+|x|^2)+3(1-|x|^2)^4 {\rm arctanh}|x|\right]\nn\\
&=&\frac{1}{(1-x^2)^4} \left[1-\frac{3}{7} x^2 +o (x^4)\right]
\equiv \frac{1}{(1-x^2)^4} [1+\alpha(x^2)]~~~.
\eea  
The relation \bref{R5dC4} is checked by the following relation
\bea
f(x^2)+\frac{2}{5}x^2\frac{d}{dx^2}f(x^2)=\frac{1}{(1-x^2)^5}~~~.
\eea
In terms of LI currents \bref{unij} the 4-form is written as
\bea
\hat{C}_{[4]}=-(1+\alpha(x^2))~{\rm tr}~
({\bf j}_{\rm unitary})\langle{\bf j}_{\rm unitary}\rangle^3  ~~~.\label{C4C4}
\eea
The 4-form potential $\hat{C}_{[4]}$ in \bref{C4C4} 
cannot be written only in terms of LI currents, but
it has explicit dependence on $x$ through  ``$\alpha(x^2)$".
Therefore we conclude that
\bea
d\hat{R}_{[5]}=0,~~\hat{R}_{[5]}=d\hat{C}_{[4]},~~
\hat{R}_{[5]}\in C^{5}(G,{\mathbb R}),~~
\hat{C}_{[4]}\notin C^{4}(G,{\mathbb R})~~~,
\eea
and the bosonic 5-form belongs to a nontrivial class of the 5-th CE cohomology group
for $G=\{GL(4)/[Sp(4)\otimes GL(1)]\}$.

Now we analyze the 5-form and 4-form by ``expansion" procedure.
This procedure realizes systematic computation not only for the leading term in the 
flat limit
but also for all order terms.
We begin with a rescaling
\bea
x^\alpha\to sx^\alpha~~,
\eea 
then LI currents are expanded as
\bea
{\bf j}_{\rm unitary}(x)\to {\bf j}_{\rm unitary}(sx)
=\displaystyle\sum_{n=1}^{\infty} s^n{\bf j}_{{\rm unitary},n}(x).
\eea 
The  MC equations \bref{mcbose} is satisfied at each power in $s$, and 
they are expanded as
\bea
\left\{
\begin{array}{rcl}
-d\langle{\bf j}_{{\rm unitary},1} \rangle&=&0\\
-d({\bf j}_{{\rm unitary},2})&=&\langle{\bf j}_{{\rm unitary},1}\rangle^2\\
-d\langle{\bf j}_{{\rm unitary},3} \rangle&=&\{({\bf j}_{{\rm unitary},2}),
\langle{\bf j}_{{\rm unitary},1}\rangle\}\\
&\vdots&
\end{array}
\right.\label{mcexpbose}~~~.
\eea
From these MC equations we read off the structure constant, 
$[T_a,T_b]=f_{ab}{}^cT_c$ $\leftrightarrow$
$dJ^c=-\frac{1}{2}f_{ab}{}^cJ^aJ^b$,
and recognize it as the structure constant
 of the
 ``expanded" algebra.
It is important that the expanded currents are no more left-invariant,
since the AdS transformations depend on $s$ as 
$\delta_{\rm AdS}=\delta_0+s\delta_1+\cdots$.  
The global Lorentz and scaling transformations 
include $\delta_0$ only
while the AdS translations includes $\delta_0$ and $\delta_1,\cdots$ in the unitary gauge. 
Expanded currents transform by themselves only under $\delta_0 x=\Lambda x$ as
${\bf j}_n~\to~h^{-1}(-\Lambda){\bf j}_n h(-\Lambda)$.

The closed 5-form and the 4-form potential are also expanded in the power of $s$ as
\bea
\hat{R}_{[5]}=\displaystyle\sum_{n=5}^{\infty} s^n \hat{R}_{[5],n}~,~~
\hat{C}_{[4]}=\displaystyle\sum_{n=5}^{\infty} s^n \hat{C}_{[4],n}~,~
\hat{R}_{[5],n}=d\hat{C}_{[4],n}~~~.
\eea
A clue of the analysis is that ``expanded" 4-form potentials $\hat{C}_{[4],n}$ 
can be 
written in terms of the ``expanded" currents ${\bf j}_n$'s.
At the lowest $s^5$ order $\hat{R}_{[5],5}$ and $\hat {C}_{[4],5}$
satisfying 
$d\hat{R}_{[5],5}=0$ and $\hat{R}_{[5],5}=d\hat {C}_{[4],5}$ are
\bea
\hat{R}_{[5],5}={\rm tr} ~\langle{\bf j}_{{\rm unitary},1}\rangle^5~
\label{r5boson5}~~,~~
\hat{C}_{[4],5}=-{\rm tr}~
 ({\bf j}_{{\rm unitary},2})\langle{\bf j}_{{\rm unitary},1}\rangle^3
 \label{r5c4boson}~~~.
\eea
This relation can be generalized for all $n>5$ as
 $\hat{R}_{[5],n}=d\hat{C}_{[4],n}$, where both $\hat{R}_{[5],n}$ 
and $\hat{C}_{[4],n}$ are written in terms of expanded currents.
Therefore it leads to that the expanded 5-forms belong to 
trivial class of the 5-th CE cohomology group
on $G'=\{$the ``expanded" $GL(4)/[Sp(4)\otimes GL(1)]$$\}$:
\bea
d\hat{R}_{[5],n}=0,~~\hat{R}_{[5],n}=d\hat{C}_{[4],n},~~
\hat{R}_{[5],n}\in C^{5}(G',{\mathbb R})~~,~~
\hat{C}_{[4],n}\in C^{4}(G',{\mathbb R})~~.\label{expCEco}
\eea

In general gauge, we introduce full $16$ variables of the $GL(4)$ matrix 
$z_m{}^a$. The $Sp(4)$ and $GL(1)$
gauge invariance removes $10+1$ degrees of freedom leaving $5$ physical
coordinates.
For example we parameterize it as 
\bea
z_m{}^a={\bf z}_m{}^a
~e^\phi e^{\varphi^{[\alpha\beta]}\gamma_{\alpha\beta}}
~~,~~
z_a{}^m=e^{-\phi}e^{-\varphi^{[\alpha\beta]} \gamma_{\alpha\beta}}~
{\bf z}_a{}^m
\label{zxzphi}
\eea
with ${\bf z}_m{}^a$ and ${\bf z}_a{}^m$ in \bref{zxz},
then the LI currents contain $\phi$ and 
$\varphi^{\alpha\beta}$ dependent terms.
The following rescaling 
\bea
\varphi^{[\alpha\beta]} ~\to~\varphi^{[\alpha\beta]} ,~~\phi ~\to~\phi,~~
x^\alpha~\to~s x^\alpha
\eea
leads to the expansion of the LI currents as
\bea
\langle{\bf j}\rangle&\to&\langle{\bf j}_1\rangle,~\langle{\bf j}_3\rangle,~
\langle{\bf j}_5\rangle,\cdots\nn\\
({\bf j})&\to&({\bf j}_0),~({\bf j}_2),~
({\bf j}_4),\cdots\label{expboson}\\
{\rm tr}~{\bf j}&\to&{\rm tr}~{\bf j}_0,~{\rm tr}~{\bf j}_2,~{\rm tr}~{\bf j}_4,~
\cdots.
\nn
\eea
They satisfy the following ``expanded" MC equations:
\bea
\left\{\begin{array}{rcl}
-d({\bf j}_0) &=&({\bf j}_0)^2
\\
-d\langle{\bf j}_1\rangle &=&\{({\bf j}_0),\langle{\bf j}_1\rangle\}
\\
-d({\bf j}_2) &=&\{({\bf j}_0),({\bf j}_2)\}
+\langle{\bf j}_1\rangle^2
\\
&\vdots&
\end{array}\right.
~~~\label{MCexp3}\label{MCexp4}
\eea 
The first relation represents the Lorentz algebra $Sp(4)$,
and all expanded currents except $({\bf j}_0)$
are transformed as bi-spinors under it.
Both the expanded 5-forms $\hat{\cal R}_{[5],n}$
and the expanded 4-forms $\hat{\cal C}_{[4],n}$
do not contain $({\bf j}_0)$ 
to guarantee the local Lorentz invariant 4-forms.  
In this gauge 
the IW contraction is easily seen by the truncation of the expanded MC equations,
because of the manifestation of the subgroup, $({\bf j}_0)$.
The IW contraction ``AdS $\to$ Lorentz" corresponds 
to the truncation of the \bref{MCexp4}
at the first level in such a way that the first equation is kept.
The IW contraction ``AdS $\to$ Poincar$\acute{\rm e}$" corresponds to the truncation 
at the second level.
The property of \bref{expCEco} is also the same in this gauge.

\par\vskip 6mm
\subsection{ Super D3-brane}

Now we will extend the bosonic analysis to the supersymmetric case.
We first find a closed and invariant 5-form in terms of the LI one forms.
The leading term includes supersymmetric version of 
the $\langle{\bf J}\rangle^5$ term \bref{r5}. 
The five form $\hat{\cal R}_{[5]}$ is obtained as
\bea
\hat{\cal R}_{[5]}&=&\s
{\rm tr}\left[
\frac{16}{15}\left( 
\langle{\bf J}\rangle^5
-\langle\bar{\bf J}\rangle^5\right)\nn\right.\\
&&~~~~+8\left(\frac{1}{3}\langle{\bf J}\rangle^3(Y)
+\langle{\bf J}\rangle^2 j_{\frac{1}{2}}\langle\bar{\bf J}\rangle
\bar{j}_{\frac{1}{2}}
-\langle\bar{\bf J}\rangle^2 \bar{j}_{\frac{1}{2}}\langle{\bf J}\rangle
{j}_{\frac{1}{2}}
-\frac{1}{3}\langle\bar{\bf J}\rangle^3(\bar{Y})
\right)\nn\\
&&~~~~\left.
+2\left(
\langle{\bf J}\rangle(Y)^2
-2\langle{\bf J}\rangle j\langle\bar{Y}\rangle\bar{j}
+2\langle\bar{\bf J}\rangle \bar{j}
\langle {Y}\rangle {j}
-\langle\bar{\bf J}\rangle(\bar{Y})^2\right)
{}^{}_{}\right], 
\label{R5S}\eea
with
\bea 
&&Y^{ab}~=~j^{a\bar{c}}\bar{j}_{\bar{c}}{}^b~~,~~\bar{Y}^{\bar{a}\bar{b}}~=~\bar{j}^{\bar{a}c}j_c{}^{\bar{b}}~~.
\eea
The closure of the 5-form $\hat{\cal R}_{[5]}$
is shown by using the MC equations \bref{MCGL44}.
It is not only closed but also manifestly invariant under 
the local
$h$ transformations \bref{htrans}.

It is important that 
\bref{R5S} can be rewritten as
\bea
\hat{\cal R}_{[5]}&=&{\cal R}_{[5]}-B_{[2]}^{NS}{\cal R}_{[3]}
\label{R5s3}\\
{\cal R}_{[5]}&=&\s
{\rm tr}\left[
\frac{16}{15}\left( \langle{\bf J}\rangle^5
-\langle\bar{\bf J}\rangle^5\right)\nn\right.\\
&&~~~~+\left. 8\left(\frac{1}{3}\langle{\bf J}\rangle^3(Y)
+\langle{\bf J}\rangle^2 j\langle\bar{\bf J}\rangle
\bar{j}
-\langle\bar{\bf J}\rangle^2 \bar{j}\langle{\bf J}\rangle
{j}
-\frac{1}{3}\langle\bar{\bf J}\rangle^3(\bar{Y})
\right)\right]
\label{R5s}
\eea
with ${\cal R}_{[3]}$ given in \bref{r3r3}. 
We take the $E=1$ gauge,
i.e. D-instanton free background, so $\hat{\cal R}_{[3]}={\cal R}_{[3]}$.
In order to write the last line of \bref{R5S} as
$-B_{[2]}^{NS}{\cal R}_{[3]}$ 
we use an identity
\bea
{\rm tr}
\left[
\langle{\bf J}\rangle (Y)^2-
2\langle{\bf J}\rangle 
j\langle \bar{Y}\rangle \bar{j}\right]
=\frac{1}{4}\left\{
{\rm tr}\left[
\langle{\bf J}\rangle \langle jj \rangle\right]~{\rm tr}\bar{j}\bar{j}
+{\rm tr}\left[
\langle\bar{\bf J}\rangle \langle \bar{j}\bar{j}\rangle\right]~{\rm tr}{j}{j}
\right\}~~~.
\eea
The last line of \bref{R5S} is combined with
${dA}~{\cal R}_{[3]}$ to give ${\cal F}~{\cal R}_{[3]}$ in \bref{generalR5} 
\bea
d{\cal L}_{WZ,{\rm AdS},{\rm D3}}&=&\hat{\cal R}_{[5]}+{dA}{\cal R}_{[3]}~
=~{\cal R}_{[5]}+{\cal F}{\cal R}_{[3]}
\label{generalR5}
\eea
and
is consistent
 with 
the result in \cite{MTD3}.

The ``super " local 4-form potential shares the same property as
the ``bosonic" one; it can be  written ``almost" in terms of LI currents
but not completely. The $\hat{\cal R}_{[5]}$ belongs to the nontrivial class of
the CE cohomology of  ${\mathcal G}$ 
\bea
d\hat{\cal R}_{[5]}=0,~~\hat{\cal R}_{[5]}=d\hat{C}_{[4]},~~
\hat{\cal R}_{[5]}\in C^{5}({\mathcal G},{\mathbb R}),~~
\hat{\cal C}_{[4]}\notin C^{4}({\mathcal G},{\mathbb R})~~~.
\eea
\medskip

Now we perform the ``expansion"  analysis 
for the supersymmetric case
to examine the properties in  \bref{expCEco}.
A coset parametrization is, for example, taken as 
\bea
Z&=&\left(\begin{array}{cc}
1+x
&\theta\\
\bar{\theta}&1+\bar{x}
\end{array}\right)
\label{cosetsuper}
\eea
where $x$ and $\bar{x}$ have $4\times 4$ components now.  
${\theta}$ and $\bar{\theta}$ are $4\times 4$ fermionic matrices. 
Rescale the coordinates as 
\bea
~x~\to~ sx,~
\theta~\to~ s^{1/2}\theta
\label{rescales}
\eea
and similar for barred variables, 
then it follows the expansion of the LI currents as
\bea
\langle{\bf J}\rangle&\to&\langle{\bf J}_1\rangle,~\langle{\bf J}_2\rangle,~
\langle{\bf J}_3\rangle,\cdots\nn\\
({\bf J})&\to&
({\bf J}_1),~
({\bf J}_2),~({\bf J}_3),\cdots\label{expsuper}\\
{\rm tr}~{\bf J}&\to&{\rm tr}~{\bf J}_1,~{\rm tr}~{\bf J}_2,~{\rm tr}~{\bf J}_3,\cdots\nn\\
j&\to&j_{\frac{1}{2}},~j_{\frac{3}{2}},~j_{\frac{5}{2}},\cdots~~.
\eea
The expanded MC equations for the super-AdS$_5\times$S$^5$ are given 
as follows:
\bea
\left\{\begin{array}{rcl}
-dj_{{\frac{1}{2}}}&=&0
\\
-d\langle{\bf J}_1\rangle&=&
\langle Y_1\rangle
\\
-d({\bf J}_1)&=&
(Y_1)
\\
-dj_{{\frac{3}{2}}}&=&
\left(\langle{\bf J}_{1}\rangle+({\bf J}_{1})\right) j_{{\frac{1}{2}}}
+j_{{\frac{1}{2}}}\left(\langle\bar{\bf J}_{1}\rangle+(\bar{\bf J}_{1})\right)
-\frac{1}{4}({\rm Str} J_1)j_{\frac{1}{2}}
\\
-d\langle{\bf J}_2\rangle&=&
\{({\bf J}_1),\langle{\bf J}_1\rangle\}
+\langle Y_2\rangle\\
-d({\bf J}_2)&=&
\langle{\bf J}_1\rangle^2
+({\bf J}_1)^2
+(Y_2)
\\
\vdots&&\end{array}\right.
\label{EMCs}
\eea
where
\bea
Y_{n}&=&\displaystyle\sum_{r<n}j_{n-r}\bar{j}_{r},~~
\left\{\begin{array}{l}
n={\rm positive~integer},\\
r={\rm positive~half~integer}
\end{array}\right.,
\eea
satisfy following MC equations
\bea
-d Y_n&=&\displaystyle\sum_{m<n}~
[{\bf J}_{n-m}, Y_{m}]~~~.
\eea

The invariant closed 5-form $\hat{\cal R}_{[5]}$  \bref{R5S}
is expanded as follows.
It starts from the $s^3$ order terms,
\bea
\hat{\cal R}_{[5],3}&=&\s {\rm tr}
\left[2
\langle{\bf J}_1\rangle(Y_1)^2
-\langle{\bf J}_1 \rangle  j_{\frac{1}{2}}\langle \bar{Y}_1\rangle 
\bar{j}_{\frac{1}{2}}
+\langle\bar{\bf J}_1 \rangle  \bar{j}_{\frac{1}{2}}\langle {Y}_1\rangle 
{j}_{\frac{1}{2}}
-2\langle\bar{\bf J}_1\rangle(\bar{Y}_1)^2
\right]\label{R53}\\
&=&-B_{[2],1}^{NS}\hat{R}_{[3]_,2}.
\nn
\eea
The corresponding  4-form potential satisfying 
$\hat{\cal R}_{[5],3}=d\hat{\cal C}_{[4],3}$ can be found 
using the MC equations \bref{EMCs}, 
\bea
\hat{\cal C}_{[4],3}&=&\s
{\rm tr}
\left[
\frac{1}{3}\left\{
2\langle{\bf J}_1\rangle \{({\bf J}_1),(Y_1)\}
+\left(\langle{\bf J}_1\rangle+({\bf J}_1)\right) j_{\frac{1}{2}}
\left(\langle\bar{\bf J}_1\rangle
+(\bar{\bf J}_1)\right)
 \bar{j}_{\frac{1}{2}}
-12 \langle{\bf J}_1\rangle j_{\frac{1}{2}}
\langle\bar{\bf J}_1\rangle \bar{j}_{\frac{1}{2}}\nn\right.\right.\\
&&~~~
-\left.\left.W_2 \left(\langle Y_1\rangle+(Y_1)\right)
\right\}
-\left\{{\rm barred~terms}\right\}\right]~~
\eea
where 4-form potentials always contain total derivative term ambiguity.
It is also convenient to define
\bea
W_n&=&j_{n-\frac{1}{2}}\bar{j}_{\frac{1}{2}}-j_{{\frac{1}{2}}}\bar{j}_{n-\frac{1}{2}}~~~,
\eea
for example $W_2=j_{{\frac{3}{2}}}\bar{j}_{\frac{1}{2}}-j_{{\frac{1}{2}}}\bar{j}_{\frac{3}{2}}$
and  it satisfies
\bea
-dW_2&=&\{{\bf J}_1,Y_1\}+2j_{\frac{1}{2}}\bar{\bf J}_1 \bar{j}_{\frac{1}{2}}~~~.
\eea

At $s^4$ order 
the 5-form and the 4-form,  satisfying
$d\hat{\cal R}_{[5],4}=0$ and $\hat{\cal R}_{[5],4}=d\hat{\cal C}_{[4],4}$,
are
\bea
\hat{\cal R}_{[5],4}&=&\s 
{\rm tr}
\left[\left\{
\frac{8}{3}
\langle{\bf J}_1\rangle^3(Y_1)
+8\langle{\bf J}_1\rangle^2 j_{\frac{1}{2}}
\langle\bar{\bf J}_1\rangle \bar{j}_{\frac{1}{2}}
+4\langle{\bf J}_1\rangle(Y_1)(Y_2)
+2\langle{\bf J}_2\rangle({Y}_1)^2 \right.\right.\nn\\
&&\left. \left.-4\left(
\langle{\bf J}_2\rangle j_{\frac{1}{2}}
\langle \bar{Y}_1\rangle \bar{j}_{\frac{1}{2}}
+\langle{\bf J}_1\rangle (j_{\frac{1}{2}}
\langle \bar{Y}_2\rangle \bar{j}_{\frac{1}{2}}
+j_{\frac{1}{2}}
\langle \bar{Y}_1\rangle \bar{j}_{\frac{3}{2}}
+j_{\frac{3}{2}}
\langle \bar{Y}_1\rangle \bar{j}_{\frac{1}{2}})
\right)
\right\}
-\left\{{\rm barred~terms}\right\}
\right]\nn\\\label{R54}
\\
\hat{\cal C}_{[4],4}&=&\s
{\rm tr}\left[\frac{1}{3}
\left\{
2\langle{\bf J}_1\rangle^3 ({\bf J}_1)
+\frac{3}{2}\langle{\bf J}_1\rangle \{({\bf J}_1),(Y_2)\}
+\frac{5}{2}\langle{\bf J}_1\rangle \{({\bf J}_2),(Y_1)\}
+2\langle{\bf J}_2\rangle \{({\bf J}_1),(Y_1)\}\right.\right.\nn\\
&&~~~\left.\left.
-\frac{1}{2}\langle Y_1\rangle [({\bf J}_2),({\bf J}_2)]
-3{\bf J}_2j_{\frac{1}{2}}\bar{\bf J}_1 \bar{j}_{\frac{1}{2}}
-12\langle{\bf J}_2\rangle
j_{\frac{1}{2}}\langle \bar{\bf J}_1\rangle \bar{j}_{\frac{1}{2}}
-{\bf J}_1\left(
j_{\frac{1}{2}}\bar{\bf J}_1 \bar{j}_{\frac{3}{2}}
+j_{\frac{3}{2}}\bar{\bf J}_1 \bar{j}_{\frac{1}{2}}
\right)\right.\right.\nn\\
&&~~~\left.\left.
+6\langle {\bf J}_1\rangle (j_{\frac{1}{2}}\langle\bar{\bf J}_{1}\rangle \bar{j}_{\frac{3}{2}}
+j_{\frac{3}{2}}\langle\bar{\bf J}_{1}\rangle \bar{j}_{\frac{1}{2}})
-\frac{1}{2}{\bf J}_1^2W_{2}
+\frac{1}{4}\left(W_3Y_1+W_2Y_2\right)
\right\}
-\left\{{\rm barred~terms}\right\}
\right]\nn\\\label{c4c4super}\label{C44}
\eea

In this way once we find the closed 5-form $\hat{\cal R}_{[5]}$, 
it is expanded as  
\bea
\hat{\cal R}_{[5]}=\displaystyle\sum_{n=3}^{\infty} s^n \hat{\cal R}_{[5],n}~~,~~
d\hat{\cal R}_{[5],n}=0~~.
\eea 
Each  $\hat{\cal R}_{[5],n}$  can be expressed as the exact form by
the ``expanded" MC equations \bref{EMCs} as
\bea
\hat{\cal R}_{[5],n}=d\hat{\cal C}_{[4],n}~~.
\eea
Then the  4-form potential is given as
\bea
\hat{\cal C}_{[4]}=\displaystyle\sum_{n=3}^{\infty} s^n \hat{\cal C}_{[4],n}~~,
~~d\hat{\cal C}_{[4]}=\hat{\cal R}_{[5]}~~.
\eea
It is important that $\hat{\cal C}_{[4]}$ can not be written 
in terms of LI currents $J$'s,
but in terms of ``expanded" currents ``$J_n$"'s.
From these facts we may deduce that the expanded 5-forms belong to 
trivial class of the 5-th CE cohomology group
on ${\mathcal G}'=$
$\{$the ``expanded" $GL(4{\mid}4)/[Sp(4)\otimes GL(1)]^2$ $\}$:
\bea
d\hat{\cal R}_{[5],n}=0,~~\hat{\cal R}_{[5],n}=d\hat{\cal C}_{[4],n},~~
\hat{\cal R}_{[5],n}\in C^{5}({\mathcal G}',{\mathbb R})~~,~~
\hat{\cal C}_{[4],n}\in C^{4}({\mathcal G}',{\mathbb R})~~.\label{SexpCEco}
\eea
As a result the WZ term for an AdS D3-brane, 
which is integration of \bref{generalR5},
is given as
\bea
{\cal L}_{WZ,{\rm AdS},{\rm D3}}=\hat{\cal C}_{[4]}+dA~\hat{\cal C}_{[2]}
\eea
with $\hat{\cal C}_{[2]}$ in \bref{C2RR}.
\par\vskip 6mm
\section{ Flat Limit}

The super-AdS algebra is reduced into the supertranslation algebra
by the IW contraction. 
In the rescaling of the coset coordinates \bref{rescales} 
the parameter $s$ plays the role of the inverse of the AdS radius,
then the flat limit corresponds to $s\to 0$ limit.

We will see how the WZ term of the D3 brane in the super-AdS$_5\times$S$^5$
background goes to one in the flat space.  
In order to get the correct flat limit of the WZ term,
the closed $(p+2)$-form $\hat{\cal R}_{[p+2]}$ 
must be reconstructed in such a way that 
 the super-AdS invariance is replaced by 
the supertranslation invariance.
The flat limit of the DBI U(1) potential $A$ must be taken suitably there.

The super-AdS variation
mixes up different order in $s$ as 
\bea
&&\delta_{\epsilon,{\rm AdS}}=\delta_{\epsilon,0}+s\delta_{\epsilon,1}\nn\\
&&\delta_{\epsilon,0}\theta^{m\bar{b}}=\epsilon^{m\bar{n}}\delta_{\bar{n}}^{\bar{b}}~,~~
\delta_{\epsilon,0}x^{ mb}=\epsilon^{ m  \bar{n}}\bar{\theta}_{\bar{n}}{}^{ b}~,~~
\delta_{\epsilon,1}\theta^{m\bar{b}}=\epsilon^{m\bar{n}}\bar{x}_{\bar{n}}{}^{\bar{b}}~~\label{51}
\eea
in the gauge \bref{cosetsuper}.
 Under the flat limit the super-AdS variation reduces to the supertranslation 
$\delta_{\epsilon,{\rm AdS}}
~~\to~~
\delta_{\epsilon,{\rm flat}}=\delta_{\epsilon,0}$.
Then expanded currents satisfies the following transformation rules
\bea
&&\delta_{\epsilon,{\rm AdS}}J=(\delta_{\epsilon,0}+s\delta_{\epsilon,1})\sum s^nJ_n
~\to~\left\{\begin{array}{l}
\delta_{\epsilon,0}j_{\frac{1}{2}}=0\\
\delta_{\epsilon,0} j_{\frac{3}{2}}+\delta_{\epsilon,1} j_{\frac{1}{2}}=0\\
\vdots
\end{array}\right.
~~
\left\{\begin{array}{l}
\delta_{\epsilon,0}J_{1}=0\\
\delta_{\epsilon,0} J_{2}+\delta_{\epsilon,1} J_{1}=0\\
\vdots
\end{array}\right.\nn\\
~~\label{expjj}
\eea
The lowest order terms in $s$ of invariant forms  are 
invariant under $\delta_{\epsilon,0}$. 
The following relations are also useful 
\bea
\left\{\begin{array}{l}
\delta_{\epsilon,0}\hat{\cal R}_{[5]3}=0\\
\delta_{\epsilon,0} \hat{\cal R}_{[5]4}+\delta_{\epsilon,1} \hat{\cal R}_{[5]3}=0\\
\vdots
\end{array}\right.
~~
\left\{\begin{array}{l}
\delta_{\epsilon,0}{\cal R}_{[3]2}=0\\
\delta_{\epsilon,0}{\cal R}_{[3]3}+\delta_{\epsilon,1}{\cal R}_{[3]2}=0\\
\vdots
\end{array}\right.\label{eR}
\eea
and 
\bea
\left\{\begin{array}{l}
\delta_{\epsilon,0}B_{[2],1}^{\rm NS}=0\\
\delta_{\epsilon,0}B_{[2],2}^{\rm NS}+\delta_{\epsilon,1}B_{[2],1}^{\rm NS}=0\\
\vdots
\end{array}\right.\label{eBB}
\eea
because $B_{[2]}$ is bilinear in fermionic currents $j$'s.

In flat space the kinetic term of the D$3$-brane action is rescaled as
$
{\cal L}_{D3}~\to~s^4{\cal L}_{D3}~~
$,
so the WZ term should also be rescaled as 
$
{\cal L}_{WZ,{\rm D3}}~\to~s^4{\cal L}_{WZ,{\rm D3}}$.
Since exterior derivative $d$ does not change the order of the expansion $n$,
the $d{\cal L}_{WZ,{\rm flat},{\rm D3}}$ is given by
\bea
d{\cal L}_{WZ,{\rm flat},{\rm D3}}&=&\displaystyle\lim_{s\to 0} \frac{1}{s^4}
d{\cal L}_{WZ,{\rm AdS},{\rm D3}}\nn\\
&=&\displaystyle\lim_{s\to 0} \frac{1}{s^4}
\left(\hat{\cal R}_{[5]}+dA~{\cal R}_{[3]}\right)\nn\\
&=&\displaystyle\lim_{s\to 0} \frac{1}{s^4}\left(
{\cal R}_{[5]}+{\cal F}~{\cal R}_{[3]}\right)\nn\\
&=&\displaystyle\lim_{s\to 0} \frac{1}{s}{\cal F}_{[2],1}{\cal R}_{[3],2}
+\left({\cal R}_{[5],4}+{\cal F}_{[2],1}~{\cal R}_{[3],3}+{\cal F}_{[2],2}~{\cal R}_{[3],2}\right).
\label{dLflat}\eea
There appear $s^3$ terms which are singular in the flat
limit.
The singular term is absent if 
\bea
{\cal F}_{[2],1}=dA_{1}-B_{[2],1}^{\rm NS}=0~~~
\label{conA1}
\eea
for $A=\displaystyle\sum_{n\geq 1} s^nA_n$. 
Since $B_{[2],1}^{\rm NS}$ is an exact form it is absorbed in 
the DBI field $A$ \cite{AdSWZ}, so that \bref{conA1} is realized.
Now the DBI U(1) field strength is
\bea
{\cal F}~=~{\cal F}_{[2],2}=dA_{2}-B_{[2],2}^{NS}~~~.
\label{conA2}
\eea
 ${\cal F}$ is superinvariant but $A_2$ and $B_{[2],2}^{NS}$ are not separately
superinvariant in the flat limit, 
contrasting to the super-AdS case in our formulation \bref{generalR}.

$\hat{\cal R}_{[5],4}$ is not supertranslation invariant from \bref{eR} and  \bref{eBB}
\bea
\delta_{\epsilon,0} \hat{\cal R}_{[5],4}&=&
-\delta_{\epsilon,1} \hat{\cal R}_{[5],3}\nn\\
&=&\left(\delta_{\epsilon,1} B_{[2],1}^{NS}\right)~\hat{\cal R}_{[3],2}
+B_{[2],1}^{NS}~\left(\delta_{\epsilon,1}\hat{\cal R}_{[3],2}\right)
\nn\\
&=&-\delta_{\epsilon,0}
\left(
B_{[2],2}^{NS}\hat{\cal R}_{[3],2}
+B_{[2],1}^{NS}\hat{\cal R}_{[3],3}
\right)~~~,
\label{Rflat}
\eea
but ${\cal R}_{[5],4}$ is supertranslation invariant 
\bea
\delta_{\epsilon,0} {\cal R}_{[5],4}&=&
\delta_{\epsilon,0} \left[\hat{\cal R}_{[5],4}-
\left(
B_{[2],2}^{NS}\hat{\cal R}_{[3],2}
+B_{[2],1}^{NS}\hat{\cal R}_{[3],3}
\right)\right]~=~0~.
\label{Rf}
\eea
The closed 5-form ${\cal R}_{[5],4}$ in a flat space is consistent with 
known results 
\cite{DpSH,KamHcano}.
The WZ term for an D3-brane
becomes
\bea
{\cal L}_{WZ,{\rm flat},{\rm D3}}=\hat{\cal C}_{[4],4}+
dA_{2}~\hat{\cal C}_{[2],2}~~~,
\eea
which is consistent with
the flat expression \cite{KamHcano}
up to total derivative terms.

The MC equations of the supertranslation algebra is obtained by
truncating the ``expanded" MC equations \bref{EMCs}
 preserving the first two equations. 
The LI currents of the supertranslation algebra are
only
 $\langle {\bf J}_1\rangle, ~j_{1/2}, ~\langle \bar{\bf J}_1\rangle,~\bar{j}_{1/2}$.
Other expanded currents are not supertranslation invariant, 
and they are rewritten in terms of the above currents with
the non-constant coefficients $x$'s and $\theta$'s.
The first term of the WZ term, $\hat{\cal C}_{[4],4}$ in \bref{c4c4super}, 
contributes to the modification of the supercharges 
producing 
 the topological D3-brane
charge in a flat space which involves the D3-brane volume $\langle dx\rangle ^3$. 
Its coefficient in the term is formally 
 $({\bf J}_1)$
  but it is not LI currents 
of the supertranslation algebra.
After gauging away the symmetric part of $dx$ from $({\bf  J}_1)$,
 it is written as 
$({\bf J}_1) \to \theta~\bar{j}_{\frac{1}{2}}$;
\bea
{\cal L}_{WZ,{\rm AdS},{\rm D3}}\sim {\rm tr}~\left[\langle {\bf J}_1\rangle^3~ ({\bf J}_1)\right]~\to~
{\cal L}_{WZ,{\rm flat},{\rm D3}}\sim 
{\rm tr}~\left[\langle {\bf J}_1\rangle^3 ~\theta~ \bar{j}_{\frac{1}{2}}\right]~~~.
\eea
Under the supertransformation, $\delta_{\epsilon,0}$,
$ {\cal L}_{WZ,{\rm AdS},{\rm D3}}$
is invariant from \bref{expjj}, but
${\cal L}_{WZ,{\rm flat},{\rm D3}}$ is not invariant any more.
There is no possibility of modification of the supercharge
involving $dx^3$  by $ {\cal L}_{WZ,{\rm AdS},{\rm D3}}$.
On the other hand there arises contribution 
to the supercharge by ${\cal L}_{WZ,{\rm flat},{\rm D3}}$,
 and we know that taking anticommutator of the supercharges 
 gives the D3-brane charge involving $dx^3$.
Since the D3-brane charge must involve $\langle dx\rangle ^3$ 
in both flat space and curved space, 
it is enough to conclude that there is no D3-brane charge
in the AdS superalgebra.
In this paper we did not consider other types of the general  topological terms 
which vanish when $\theta\to 0$ \cite{Peeters}.

\par\vskip 6mm
\section{ Discussion and conclusions}

We have obtained  concrete expression of the WZ terms 
for D$p$-branes with $p=-1,1,3$ in the AdS$_5\times$S$^5$ space.
D-instanton is described by the auxiliary degree of freedom of $GL(1)$
in the Roiban-Siegel formalism.
We have shown that the WZ term for a D3-brane in the AdS space
can not be written in terms of the LI currents but in terms of the
``expanded" currents.
   This ``expansion" procedure can be generalized as the prescription 
of obtaining an expression of a
$(p+1)$-form WZ term for a $p$-brane in AdS spaces.

Flat limit of the WZ term for a D3-brane is also examined. 
 Since the super-AdS transformation rules depend on
the expansion parameter,
the supertranslation invariance and the closure of the WZ 
term must be examined carefully in the flat limit.
The degree of freedom of the DBI field plays essential role 
to absorbe the divergent term in the flat limit.

The ``would-be" topological term, $dx^3$, 
in the flat WZ term
can be traced back into the AdS WZ term
as tr$\langle {\bf J}_1\rangle^3({\bf J}_1)$.
This term keeps the left-invariance ($\delta_{\epsilon,0}$) in the AdS space,
so the corresponding  term does not modify 
the supercharges in the AdS space.
There is no  topological extension of the 
super-AdS algebra.
The indices of the ``would-be" topological term, 
$\langle {\bf J}\rangle^3_{(ab)}$,
has the same indices as  the Lorentz rotation 
generator
$M_{(ab)}$, so 
it can be absorbed into the  Lorentz rotation 
generator
analogous to \cite{MHD0}.

For a fundamental superstring (F1) case the AdS-WZ term is given by
${\cal L}_{WZ,{\rm AdS},{\rm F1}}=B^{NS}_{[2]}$ as \bref{31}.  
In flat space the action of the F$1$-brane is rescaled as
$
{\cal L}_{F1}~\to~s^2{\cal L}_{F1}~~
$;
\bea
{\cal L}_{WZ,{\rm flat},{\rm F1}}&=&\displaystyle\lim_{s\to 0} \frac{1}{s^2}
{\cal L}_{WZ,{\rm AdS},{\rm F1}}\nn\\
&=&\displaystyle\lim_{s\to 0} \frac{1}{s}B^{NS}_{[2],1}
+B^{NS}_{[2],2}~~~,
\label{Lflat}
\eea
so the singular term $B^{NS}_{[2],1}$ must be subtracted in the limiting procedure.
The non-singular term $B^{NS}_{[2],2}$ corresponds to the familiar WZ term for a flat superstring.
Since $B^{NS}_{[2],1}$ is a total derivative term, it may be also subtracted 
from the WZ term for a superstring in the AdS space
\bea
\tilde{\cal L}_{WZ,{\rm AdS},{\rm F1}}=B^{NS}_{[2]}-B^{NS}_{[2],1}=\displaystyle\int_0^1
dt~ H^{NS}_{[3]}(t)\label{MTWZ}
\eea
which corresponds to the one given by Metsaev and Tseytlin \cite{MT}.
As shown in \bref{eBB} $B_{[2],1}^{NS}$ is not super-AdS invariant 
 but pseudoinvariant,
$\tilde{\cal L}_{WZ,{\rm AdS},{\rm F1}}$ is also pseudoinvariant under the 
super-AdS transformation.
In this case we have a string charge in the anticommutator 
of supercharges which corresponds to the one calculated in \cite{AdSWZ}.
\footnote{We thank Peeters and Zamaklar for suggesting the following consideration.
In our notation the subtracted total derivative term is $B_{[2],1}^{NS}=\frac{1}{2}{\rm tr}
\left(d\theta d\theta-d\bar{\theta}d\bar{\theta}\right)$.
Under the super-AdS transformation \bref{51} in this gauge 
the WZ term becomes total derivative term as
\bea
\delta_\epsilon\tilde{\cal L}_{WZ,{\rm AdS},{\rm F1}}=d ~{\rm tr}
\left(d \delta_\epsilon \theta ~\theta-d \delta_\epsilon\bar{\theta}~\bar{\theta}\right)
\label{fn1}
\eea
where $dx$ is preferred than bare $x$ at the time-component surface term. 
We denote the global GL(4${\mid}$4) generators as $G_{MN}=Z_M{}^A\Pi_{AN}$
with canonical conjugates $\Pi_A{}^M$.
These charges get total derivative term contribution from \bref{fn1} as demonstrated in \cite{MHD0}
\bea
\tilde{G}_{m\bar{n}}=G_{m\bar{n}}+\displaystyle\int(\partial_\sigma x_{m}{}^{a})\bar{\theta}_{\bar{n}a}~~,~~
\tilde{\bar{G}}_{\bar{m}n}=\bar{G}_{\bar{m}n}-\displaystyle\int(\partial_\sigma \bar{x}_{\bar{m}}{}^{\bar{a}})\theta_{n\bar{a}}
~~,
\eea
then anticommutator of $\tilde{G}_{m\bar{n}}$'s produces the string charge as
\bea
\{\tilde{G}_{m\bar{n}},\tilde{G}_{l\bar{k}}\}=\Omega_{\bar{n}\bar{k}}
\displaystyle\int\partial_\sigma
\left(x_m{}^a x_{la}\right)~~~
\eea
which corresponds to the string charge in \cite{AdSWZ}.
 }
This superalgebra containing the string charge satisfies the Jacobi identity. 
This formulation manifests the topological term in the flat limit.
In the AdS space it seems that  
both ${\cal L}_{WZ,{\rm AdS},F1}$ in \bref{31} and 
$\tilde{\cal L}_{WZ,{\rm AdS},F1}$ in \bref{MTWZ}
are allowed.

The point is that 
WZ terms for branes in the flat space can not be superinvariant resulting the brane charges
in their superalgebras,
but the ones in the AdS space can be ``superinvariant" resulting no brane charges.
Careful flat limiting leads to the correct flat BPS condition realized
as the topological extension of the superalgebra.

\par\vskip 6mm

\noindent{\bf Acknowledgments}

M.H. thanks to Shunya Mizoguchi for fruitful discussions especially 
on cohomology.


\end{document}